\documentclass[prc,preprint,amsmath,amssymb,
floatfix]{revtex4-1}
\usepackage[utf8]{inputenc}
\usepackage{amsfonts}
\usepackage{graphicx}
\usepackage{fullpage}
\usepackage{mathtools}
\usepackage{bm}
\usepackage{subfig}

\usepackage{CJKutf8}
\usepackage{mathrsfs}
\usepackage{bbm}
\usepackage{graphicx}
\usepackage{dcolumn}
\usepackage{bm}
\usepackage{float}
\usepackage{color}
\usepackage{multirow}
\usepackage{enumerate}
\usepackage{BOONDOX-cal}
\usepackage{listings}
\usepackage{xcolor}

\begin{document}
%\begin{CJK*}{GB}{} % Use default fonts from CJK (see below)

%\begin{CJK*}{UTF8}{}
%\CJKfamily{gbsn}
%\begin{frontmatter}
\title{ Shannon entropy of optimized proton-neutron pair condensates }

\author{ Shu-Yuan Liang }
\affiliation{College of Physics and Engineering, Qufu Normal University, 57 Jingxuan West Road, Qufu, Shandong 273165, China}

\author{Yi Lu }
\email{luyi@qfnu.edu.cn}
\affiliation{College of Physics and Engineering, Qufu Normal University, 57 Jingxuan West Road, Qufu, Shandong 273165, China}

\author{Yang Lei }
\email{leiyang19850228@gmail.com}
\affiliation{School of National Defense Science and Technology, Southwest University of Science and Technology, Mianyang 621010, China}

\author{Calvin W. Johnson }
\email{cjohnson@sdsu.edu}
\affiliation{ Department of Physics, San Diego State University, 5500 Campanile Drive, San Diego, CA 92182, USA}

\author{Guan-Jian Fu}
\email{gjfu@tongji.edu.cn}
\affiliation{ School of Physics Science and Engineering, Tongji University, Shanghai 200092, China}

\author{Jia Jie Shen}
\affiliation{ School of Arts and Sciences, Shanghai Maritime University, Shanghai 201306, China }

\begin{abstract}
Proton-neutron pairing and like-nucleon pairing are two different facets of atomic nuclear configurations.
While like-nucleon pair condensates manifest their superfluidic nature in semi magic nuclei, it is not absolutely clear if there exists a T=0 proton-neutron pair condensate phase in $N=Z$ nuclei.
With an explicit formalism of general pair condensates with good particle numbers, we optimize proton-neutron pair condensates for all $N=Z$ nuclei between $^{16}$O and $^{100}$Sn, given shell model effective interactions.
As comparison, we also optimize like-nucleon pair condensates for their semi-magic isotones.
Shannon entanglement entropy is a measurement of mixing among pair configurations, and can signal intrinsic phase transition.
It turns out the like-nucleon pair condensates for semi-magic nuclei have large entropies signaling an entangled phase, but the proton-neutron pair condensates end up not far from a Hartree-Fock solution, with small entropy.
With artificial pairing interaction strengths, we show that the general proton-neutron pair condensate can transit from an entangled T=1 phase to an entangled T=0 phase, i.e. pairing phase transition driven by external parameters.
In the T=0 limit, the proton-neutron pair condensate optimized for $^{24}$Mg turns out to be a purely P pair condensate with large entanglement entropy, although such cases may occur in cold atom systems, unlikely in atomic nuclei.
\end{abstract}
	
\maketitle

\section{Introduction}\label{sec-int}
Pairing is a fundamental feature of many-body systems ranging from superconducting electronic systems, ultracold atoms, to atomic nuclei.
The atomic nuclei have a unique two-partite composition, as they are made of two types of fermions, i.e. protons and neutrons, and the interactions among them approximately respect isospin symmetry.
T=1 pairing is symmetric for proton-proton, neutron-neutron and proton-neutron pairs, while T=0 pairing is only for proton-neutron pairs.
The T=1 pairing of like-nucleons \cite{Bohr1958} has important influence in nuclear binding energies, isotopic abundances, moment of inertia, and semi-magic nuclear spectrum.
In semi magic nuclei, the general seniority scheme can successfully describe energetic spectrum and electromagnetic transition strengths\cite{1981Simple},
and it is believed that the ground state of a semi magic nucleus can be well approximated with a T=1 $S$-pair (angular momentum $J=0$) condensate, or the seniority-0 state. 
The breaking of $S$ pairs, or higher-seniority configurations, explain the low-lying excited states.
The T=1 $S$-pair condensate has a superconducting nature similar to the BCS phase in condensed matter physics;
if there exists a T=0 correlated proton-neutron pair condensate phase in the $N=Z$ nuclei, it would be an interesting feature of atomic nuclei the two-partite system.

In 2010, a spin-aligned phase\cite{Cederwall2010} (or a stretch scheme \cite{Danos1967}), was suggested to explain experimental data of the $N=Z$ nucleus $^{92}$Pd, considering four $J=9$ pairs of proton-neutron holes condensate in the ground state.
Based on expectation of T=0 pair counting operators with shell model wavefunctions, it is suggested the yrast states are predominantly made of $(0g_{9/2})^2_{J=9}$ proton-neutron pairs\cite{Qi2011}.
When two $J=9$ proton-neutron stretched pairs couple into total angular momentum zero, it is partly equivalent to coupling proton low-angular-momentum pairs with neutron ones\cite{Zerguine2011,Neergaard2013,Fu2013}. 
Possible quartet condensation is also an interesting degree of freedom \cite{PhysRevC.91.054318,Sandulescu2015}.
There has been extensive discussion, see Ref. \cite{FRAUENDORF201424} and references therein.

Based on Hartree-Fock-Bogolyubov calculations with schematic interactions, recently it is suggested spin-triplet condensates (T=0) prevail in nuclei with $N = Z \approx 65$\cite{PhysRevLett.106.252502}. Although these nuclei are beyond the proton drip line, this effect can lead to mixed-spin character in $N > Z \approx 65$ nuclei beneath the proton drip line, while other nuclei show a purely spin-singlet character (T=1)\cite{PhysRevC.93.014312}.
Studies based on boson mapping techniques have shown interesting analytic features of P pair (T=0, J=1) condensates\cite{VanIsacker2016}.
Ultracold atoms can be manipulated with lasers to separate spin phases\cite{Lin2011}, adding interest and significance to these discussions.
However realistic calculations based on general proton-neutron pair condensates with good particle numbers are yet scarce\cite{Romero2019}.

In this work we use a general proton-neutron pair condensate as the trial wavefunction, 
\begin{equation}
|\pi\nu; PC\rangle  \equiv (\hat{A}^\dagger_{\pi \nu})^n | 0 \rangle,~~~~ \hat{A}^{\dagger}_{\pi\nu} \equiv \frac{1}{2}\sum_{\alpha \in \pi, \beta \in \nu} A_{\alpha \beta} \hat{c}^{\dagger}_\alpha \hat{c}^{\dagger}_\beta,
\label{eqn-pnpc}
\end{equation}
where $\alpha, \beta$ stands for single-particle orbits, and $A_{\alpha \beta}$ are the pair structure coefficients to be determined.
We optimize $A_{\alpha\beta}$ variationally, so as to minimize the expectation of shell model effective interactions,
\begin{equation}
\min_{A_{\alpha\beta}} \frac{ \langle \pi\nu; PC | \hat{H} | \pi\nu; PC \rangle }{ \langle \pi\nu; PC | \pi\nu; PC \rangle }.
\end{equation}
Then we implement triaxial angular momentum projection to restore rotational symmetry and approximate nuclear wavefunctions with good quantum numbers.
Variation after angular momentum projection (VAP) leads to lower energies but are more complicated and time-consuming; hence we restrict ourselves to projection after variation (PAV).
Similar approaches with angular momentum projections include the Projected Hartree-Fock method \cite{PhysRevC.66.034301,Johnsonb}, the Projected Hartree-Fock-Bogolyubov method with angular momentum projection
 \cite{Ring1980,PhysRevC.100.044308,PhysRevC.104.054306,PhysRevC.103.024315,PhysRevC.103.014312,R.Rodriguez-Guzman2002}, the Projected Shell Model \cite{PhysRevC.61.064323} and so on.

To analyze theoretical wavefunctions for a possible proton-neutron pair condensate phase, it was suggested the invariant correlation entropy can be used as a signal\cite{FRAUENDORF201424}.
The invariant correlation entropy\cite{Sokolov1998} is a von Neumann entropy, with the virtue that it is basis independent, and can be used to measure the sensitivity of a state to external parameters\cite{Volya2003}.
Indeed the quantum entanglement entropy has been a sourceful idea to measure many-body wavefunctions.
Orbital entanglement entropy has been used in ab initio calculations to analyze the efficacy of different bases\cite{Robin2021}, and has been used in shell model computations to reveal small entanglement between proton and neutron sectors in neutron rich heavy nuclei\cite{Johnson2023}, motivating new truncation strategies\cite{Gorton2024}. 
The basis-dependent Shannon entropy has also been used to detect competence between single particle behavior and pairing forces, leading to possible pairing phase transition in rare earth nuclei\cite{Guan2016}.

In this work we consider the Shannon entropy as a measurement of the pairing entanglement of the optimized proton-neutron pair condensates, and analyze the entropy as a function of T=0 and T=1 pairing forces.
A mean field solution, i.e. an optimized Slater determinant at an energetically lowest minimum, has zero entropy, while in ideal conditions, the pair condensate solution can have maximum entanglement entropy.
Therefore, the entropy is a diagnostic signal of a possible entangled ``phase" of the proton-neutron pair condensate.

In Sec. \ref{formalism} we introduce the formalism of projection after variation of a general proton-neutron pair condensate, as well as the definition of entanglement entropy we use in this work; in Sec. \ref{results} we show the spectrum and B(E2) values from one proton-neutron pair condensate, and present systematic results of $N=Z$ nuclei from $^{18}$F to $^{98}$In, as well as pairing ``phase" transition with artificial interactions; in Sec. \ref{summary} we summarize this work and discuss the outlooks.

\section{Formalism}\label{formalism}
\subsection{Interactions and Configurations}
In the framework of the shell model, we use widely accepted effective interactions, i.e. USDB\cite{PhysRevC.74.034315} in the $1s0d_{3/2}0d_{5/2}$ major shell, GX1A\cite{PhysRevC.65.061301} in the $1p_{1/2}1p_{3/2}0f_{5/2}0f_{7/2}$ major shell, JUN45 \cite{Honma2009} in the $1p_{1/2}1p_{3/2}0f_{5/2}0g_{9/2}$ major shell, with 1- and 2-body parts in the occupation space,
\begin{equation}
\begin{aligned}
\hat{H} = & \hat{H}_0 + \hat{H}_{T=0} + \hat{H}_{T=1} \\
& = \sum_{a} \epsilon_a \hat{n}_a 
+ \sum_{abcd } \frac{\sqrt{(1+\delta_{ab})(1+\delta_{cd})}}{4} 
\sum_{IT} V(abcd;IT) \sum_{M M_T} \hat{A}^\dagger_{IM, T M_T}(ab) \hat{A}_{IM, T M_T}(cd),
\end{aligned}
\label{eqn-hamiltonian}
\end{equation}
where $\hat{A}^\dagger_{IM, TM_T}(ab)$ and  $\hat{A}_{I M, TM_T}(cd)$ are so-called ``non-collective" pair creations and annihilation operators: 
$\hat{A}^\dagger_{IM, TM_T}(ab) = \left [\hat{c}^\dagger_a \otimes \hat{c}^\dagger_b \right ]_{IM, T M_T},$ where $\otimes$ signals coupling via Clebsch-Gordan coefficients, 
and $\hat{A}_{IM, T M_T}(ab) = \left ( \hat{A}^\dagger_{IM, TM_T}(ab) \right)^\dagger$. 
These effective interactions have been adjusted to hundreds of experimental data for energetic spectrum and electromagnetic transition rates \cite{PhysRevC.74.034315, PhysRevC.65.061301, Honma2009}.
%With effective interactions and the valence space given, the shell model consider all the possible configurations and diagonalize the Hamiltonian matrix exactly.
%However the configuration space dimension increases exponentially with the valence particle numbers, therefore the shell model computations are almost trivial in the $1s_{1/2}0d_{3/2}0d_{5/2}$ major shell, approachable in the $1p_{1/2}1p_{3/2}0f_{5/2}0f_{7/2}$ major shell with super computers, but becomes intractable for most rotational nuclei in the $2s_{1/2}1d_{3/2}1d_{5/2}0g_{7/2}0h_{11/2}$ major shell. 
%Therefore we resort to variational methods, and implement projection after variation, so as to restore good quantum numbers of the angular momentum.

A general pair condensate is defined as 
\begin{equation}
(\hat{A}^\dagger)^n | 0 \rangle,~~~~ \hat{A}^{\dagger}=\frac{1}{2}\sum_{\alpha \beta} A_{\alpha \beta} \hat{c}^{\dagger}_\alpha \hat{c}^{\dagger}_\beta,
\label{eqn-pair-condensate}
\end{equation}
where $|0\rangle$ is the bare fermion vacuum, and $\hat{c}^{\dagger}_\alpha, \hat{c}^{\dagger}_\beta$ are fermion creation operators with $\alpha, \beta$ labelling single-particle states with quantum numbers $(n l j m)$, which can be (though not restricted to) spherical harmonic bases of the shell model, and, finally, $A_{\alpha \beta}$ are the ``pair structure coefficients".
Without losing generality, $A_{\alpha \beta}$ makes a skew complex matrix, though in practice $A_{\alpha \beta}$ is restricted to be real.
A Slater determinant of the Hartree-Fock method, a seniority-0 state, or a number projected component of the BCS/HFB vacuum are all special cases of the general pair condensate as in Eq. (\ref{eqn-pair-condensate}), therefore this trial wavefunction reserves more generality\cite{PhysRevC.44.R598}.
In quantum chemical physics, a general pair condensate is also known as ``geminal wavefunction" \cite{Moisset2022}.

\subsection{Variation} \label{subsec:variation}
Variational formulas of like-nucleon pair condensates has been proposed in different forms \cite{PhysRevC.44.R598,PhysRevC.26.2640,mcsm-pair,PhysRevC.102.024310,PhysRevC.105.034317}, and recently triaxially angular momentum projected states out of a like-nucleon pair condensate has been benchmarked in shell model spaces with effective interactions\cite{PhysRevC.105.034317}.
In this work we extend the formalism in Ref. \cite{PhysRevC.105.034317} from like-nucleon pair condensates to proton-neutron pair condensates.
We assume a general proton-neutron pair condensate as the trial wavefunction ansatz,
$|\pi\nu; PC\rangle  = (\frac{1}{2}\sum_{\alpha \in \pi, \beta \in \nu} A_{\alpha \beta} \hat{c}^{\dagger}_\alpha \hat{c}^{\dagger}_\beta)^n | 0 \rangle$, as defined in Eq. (\ref{eqn-pnpc}).
The pair coefficients $A_{\alpha \beta}$ are parameters to be determined by variation, so to minimize the energy
$\min_{A_{\alpha\beta}}  \langle \pi\nu; PC | \hat{H} | \pi\nu; PC \rangle / \langle \pi\nu; PC | \pi\nu; PC \rangle $.
One can restrict $A_{\alpha \beta}$ to good isospin a priori, by demanding
\begin{equation}
A_{b \pi, a \nu} = (-1)^{T} A_{a \pi, b \nu}, ~~~~T = 0,1,
\end{equation}
where $a, b$ stands for Harmonic Oscillator orbits with good quantum numbers $(nljm)$,
but we allow it to vary generally, therefore it is a general $T_z = 0$ pair, with both T=0 and T=1 components.
We use the explicit formalism presented in Ref. \cite{PhysRevC.105.034317} to do variation, until an energetic minimum is reached.

\begin{figure}[ht!] 
	\centering\includegraphics[width=0.7\textwidth]{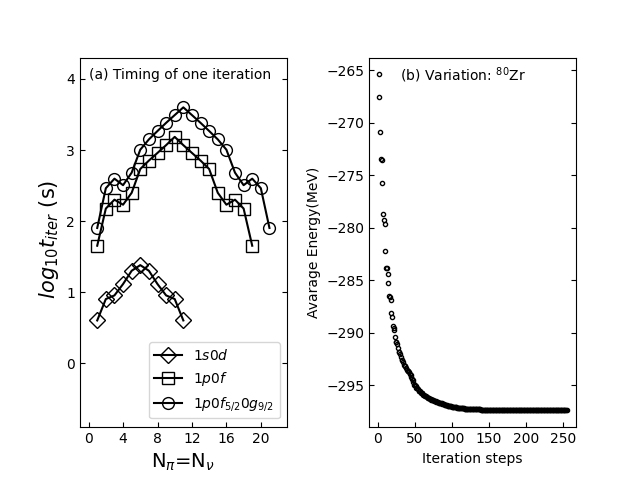}
	\caption{ Panel (a): Logarithm CPU time (s) of one iteration in the variation of one proton-neutron pair condensate for $N=Z$ nuclei, using conjugate gradient method wrapped in the GNU Scientific Library function gsl\_multimin\_fdf\_minimizer \cite{gsl}. All these data are taken on a 48-core workstation. Panel (b): Energy descent in the variation of one proton-neutron pair condensate for $^{80}$Zr in the $1p0f_{5/2}0g_{9/2}$ major shell. }
\label{fig:timing}
\end{figure}

In Fig. \ref{fig:timing} the logarithm total CPU time (s) of one iteration is recorded in the variation of one proton-neutron pair condensate for $N=Z$ nuclei, in $1s0d, 1p0f, 1p0f_{5/2}0g_{9/2}$ major shells.
There are 144 free parameters for $1s0d$ shell, 400 for $1p0f$,  and 484 for $1p0f_{5/2}0g_{9/2}$, respectively.
The total CPU time is collected on a 48-core workstation.
As pointed out in Ref. \cite{PhysRevC.105.034317}, the complexity is polynomial.
After half the major shell is filled, the Pandya particle-hole transformation is activated and we work with holes instead of particles, therefore the CPU time peaks at half shells.
We also show an variation example of $^{80}$Zr in the $1p0f_{5/2}0g_{9/2}$ major shell.
The average energy, or the expectation of the Hamiltonian on a proton-neutron pair condensate, steeply descend in the first 20 iterations, and then saturates after about 200 iterations.
After 200 iterations, the further descent of the energy is negligible.
The iteration ends when the module of the gradient vector is smaller than a preset cutoff value.

\begin{figure}[ht!] 
	\centering\includegraphics[width=0.7\textwidth]{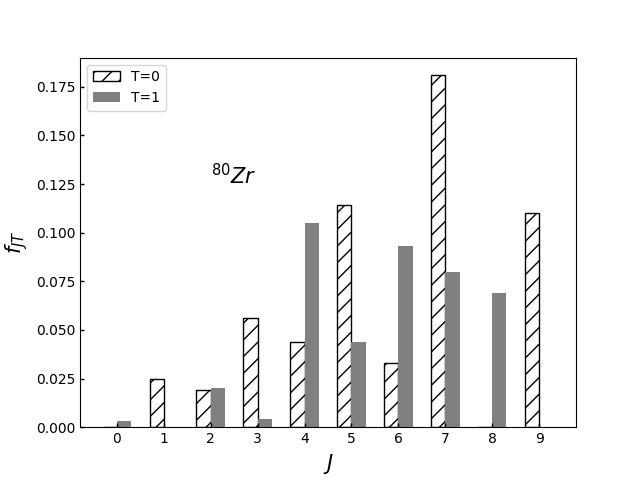}
	\caption{ Decomposition of the optimized proton-neutron pair for $^{80}$Zr in the $1p0f_{5/2}0g_{9/2}$ major shell. }
\label{fig:pair_decomposition}
\end{figure}

The optimized proton-neutron pair can be decomposed into components with good quantum numbers $(JMT)$,
\begin{equation}
\hat{A}^\dagger_{\pi \nu} = \frac{1}{2}\sum_{\alpha\in\pi, \beta \in \nu} A_{\alpha \beta} \hat{c}^\dagger_\alpha \hat{c}^\dagger_\beta
= \sum_{JMT} \hat{\mathbbm A}^{JMT\dagger}_{\pi \nu}
%= \sum_{JM} {\mathcal A}^{JM\dagger}_{\pi \nu},
\end{equation}
where $\hat{\mathbbm A}^{JMT\dagger}_{\pi \nu}$ is a pair with good angular momentum quantum numbers $(JM)$ and good isospin $T=0,1$,
\begin{eqnarray}
\hat{\mathbbm A}^{JMT\dagger}_{\pi \nu} = \sum_{a b} y(abJMT) (\hat{c}^\dagger_a \otimes \hat{c}^\dagger_b)_{JMT}.
\end{eqnarray}
The ``structure coefficients" $y(abJMT)$ of the $(JMT)$ component $\hat{\mathbbm A}^{JMT\dagger}_{\pi \nu}$ can be derived,
\begin{eqnarray}
y(abJM T) &=& [(-1)^{1+T} z(abJM) - (-1)^{j_a + j_b + J} z(baJM) ]/ \sqrt{2}, \\
z(abJM) &=& \frac{1}{2}\sum_{m_a m_b} A_{j_a m_a, j_b m_b} (j_a m_a j_b m_b|JM),
\end{eqnarray}
where $(j_a m_a j_b m_b|JM)$ is the Clebsch-Gorden coefficient.
The fraction of components with $(JT)$ in the optimized proton-neutron pair $\hat{A}^\dagger_{\pi\nu}$ sums over $M$,
\begin{equation} \label{eqn:fJT}
f_{JT} \equiv \sum_{M} \frac{\langle \hat{\mathbbm A}_{JMT} | \hat{\mathbbm A}^\dagger_{JMT} \rangle  }{\langle \hat{A}_{\pi \nu}  | \hat{A}^\dagger_{\pi \nu}  \rangle}.
\end{equation}
In Fig. \ref{fig:pair_decomposition} we present the decomposition of the optimized proton-neutron pair for $^{80}$Zr in the $1p0f_{5/2}0g_{9/2}$ shell.
With M-scheme shell model codes. e.g. the BIGSTICK \cite{2013Factorization}, $^{80}$Zr has configuration dimension at the order of $10^{10}$, and therefore is not easily accessible.
Generally speaking, T=0 odd-$J$ components, and T=1 even-$J$ components are in dominance in the optimized proton-neutron pair condensate.

\subsection{Canonical basis and Shannon entropy}
In matrix theory, it was proved that any given skew matrix $A_{\alpha \beta}$ can be transformed into a canonical form \cite{Zumino1962},
\begin{equation}\label{A=UXUtop}
A = UXU^\top,
\end{equation}
where $U$ is a unitary matrix, and $X$ is block diagonal $\left\{ [ \begin{smallmatrix}
0 & \nu_1 \\ -\nu_1 & 0 
\end{smallmatrix}] ,  
[ \begin{smallmatrix}
0 & \nu_2 \\ -\nu_2 & 0 
\end{smallmatrix}] ,
\cdots
\right\}$, with $\nu_k >0$.
This means a general pair $\hat{A}^\dagger$ can be always rewritten into a canonical form,
\begin{equation}\label{canonical-basis}
\hat{A}^{\dagger}=\frac{1}{2}\sum_{\alpha \beta} A_{\alpha \beta} \hat{c}^{\dagger}_\alpha \hat{c}^{\dagger}_\beta
=  \sum_{k>0} \nu_k a^\dagger_k a^\dagger_{\bar{k}},
\end{equation}
where $a^\dagger_k \equiv \sum_\alpha U_{\alpha k} \hat{c}^\dagger_\alpha$.
This is reminiscent of the Bloch-Messiah theorem in the HFB formalism \cite{Ring1980}, where Bogolyubov quasiparticles are built upon the canonical basis, therefore we denote Eq. (\ref{canonical-basis}) as the ``canonical transformation" of a general pair, and $a^\dagger_k, a^\dagger_{\bar{k}}$ as ``canonical basis" on which a seniority-type pair structure naturally shows up.
It is also linked to Kramer's degeneracy when $(k,\bar{k})$ are time-reversal pairs\cite{Yu2022}.
In the limit of a Slater determinant with $2n$ particles, 
\begin{equation}
\nu_k \left\{
\begin{aligned}
&\neq 0, &k = 1,\cdots, n;\\
&= 0, ~~~ &k = n+1, \cdots, \Omega, 
\end{aligned}
\right.
\end{equation}
where $\Omega$ is the occupancy of pairs, e.g. $\Omega=16$ in the $50-82$ major shell.
In the limit of uniform pairing, i.e. $\nu_k = \frac{1}{\sqrt{\Omega}}$, a normalized pair condensate with $n$ pairs is
\begin{equation}\label{pair-condensate}
|PC, n\rangle = \sqrt{\frac{(\Omega -n)!\Omega^n}{\Omega! n!}} (\hat{A}^\dagger)^n | 0\rangle.
\end{equation}
Adding one new pair, it results in
\begin{equation}
\hat{A}^\dagger |PC, n\rangle = \sqrt{\frac{(n+1)(\Omega-n)}{\Omega}} |PC, n+1\rangle.
\end{equation}
When $\Omega \rightarrow \infty$, this is reminiscent of 1-d harmonic oscillator $\hat{a}^\dagger |n\rangle = \sqrt{n+1}|n+1\rangle$ in quantum mechanics textbooks.
In this sense, it can be considered as an ``entangled pair-condensate", different from a mean-field Slater determinant.

As the canonical basis in (\ref{canonical-basis}) is unique given the pair condensate (\ref{pair-condensate}), we use the Shannon entropy, though we believe using the invariant correlation entropy \cite{Volya2003} leads to the same physical conclusions.
We define the Shannon entropy as \cite{Guan2016}
\begin{equation}
S = \sum_{\{k_1, k_2, \cdots, k_n\}} - p(k_1 k_2 \cdots k_n)~ log_d~ p(k_1 k_2 \cdots k_n), \label{eqn:entropy}
\end{equation}
where $p(k_1 k_2 \cdots k_n)$ is the probability of finding the configuration $a^\dagger_{k_1} a^\dagger_{\bar{k}_1} a^\dagger_{k_2} a^\dagger_{\bar{k}_2} \cdots a^\dagger_{k_n} a^\dagger_{\bar{k}_n}$ in the pair condensate $|PC, n\rangle$,
\begin{equation}
p(k_1 k_2 \cdots k_n) = \frac{ \nu^2_{k_1} \nu^2_{k_2}\cdots \nu^2_{k_n} }{ \langle PC, n|PC, n\rangle },
\end{equation}
where $\nu_k$, as defined in (\ref{canonical-basis}), is the amplitude of non-collective pair $a^\dagger_k a^\dagger_{\bar{k}}$ in the collective pair $\hat{A}^\dagger$.
The Shannon entropy $S$ quantifies how close a general pair condensate in Eq. (\ref{eqn-pair-condensate}) is to a Slater determinant, 
where $\{k_1, k_2, \cdots, k_n\}$ traverse all $C^n_\Omega$ different choices of $n$ different $(k, \bar{k})$ pairs out of all $\Omega$ ones, i.e. all the pairing configurations.
The logarithmic base $d$ equals $C_\Omega^n$, to scale $S$ such that $S \in [0,1]$ for any given particle number.
In the limit of a Slater determinant, only one $p(k_1 \cdots k_n)$ value equals $1$ and others equal $0$, and correspondingly the entropy $S=0$, signaling a mean field solution;
in the limit of uniform pairing, $S$ reaches its maximum value $S=1$, signaling the ``entangled phase".

\subsection{Triaxial angular momentum projection}

The triaxial angular momentum projection via an linear algebraic approach \cite{PhysRevC.96.064304} is implemented, to restore good angular momentum of the proton-neutron pair condensate (\ref{pn-pc}).
When the maximum angular momentum is $10\sim20 \hbar$, the linear algebraic approach is about 10 times faster than the traditional quadrature method \cite{2019Convergence}.

Given a proton-neutron pair condensate, and an angular momentum quantum number $J$, there are at most $2J+1$ projected bases,
\begin{equation}
\left\{ \hat{P}^J_{M,-J} | \pi \nu; PC \rangle, ~~ \hat{P}^J_{M,-J+1} | \pi \nu; PC \rangle,~~ \cdots,~~ \hat{P}^J_{M,J} | \pi \nu; PC \rangle \right\}.
\end{equation}
In order to approximately diagonalize the Hamiltonian in the linear space of the projected bases, we compute the kernel values, i.e. the inner products of projected bases and Hamiltonian matrix elements,
\begin{eqnarray}
{\mathcal N}^J_{MK}  \equiv  \langle PC;\pi\nu | \hat{P}^J_{MK} | PC;\pi\nu \rangle , ~~~
{\mathcal H}^J_{MK}  \equiv  \langle PC;\pi\nu | \hat{H} \hat{P}^J_{MK} | PC;\pi\nu \rangle , 
\end{eqnarray}
and solve the Hill-Wheeler equation,
\begin{eqnarray}
\sum_K {\mathcal H}^J_{K'K} g^r_{JK} = \epsilon_{r, J} \sum_K {\mathcal N}^J_{K' K} g^r_{JK}. \label{eqn:HillWheeler}
\end{eqnarray}
Thus configuration mixing is introduced among projected bases.
$\epsilon_{r,J}$ denotes the energy of the $r$-th eigenstate with angular momentum $J$.
$g^r_{JK}$ values determine the projected wavefunction.
With the projected wavefunctions, electromagnetic transition strengths can be further computed\cite{PhysRevC.105.034317}.

\section{Results and Discussions}\label{results}

In Subsection \ref{subsection:energy+BE2} we present one-reference results, i.e. energetic spectrum and B(E2) results generated from one ``intrinsic" proton-neutron pair condensate, so to benchmark yrast states projected from a PN-PC, against the shell model and the Projected Hartree-Fock method.
In Subsection \ref{subsec:shannon-entropy} we present systematically the Shannon entropy of optimized proton-neutron pair condensates for $N=Z$ nuclei from $^{18}$F to $^{98}$In, in hope to discuss the possibility of an entangled proton-neutron pair condensate phase in $N=Z$ nuclei.
In Subsection \ref{subsec:t1t0-competition} further analysis is implemented for possible phase transition of the optimized proton-neutron pair condensate from the T=1 limit to the T=0 limit, by artificially adjusting the external parameters in the Hamiltonian.

\subsection{Energetic spectrum and B(E2) values}\label{subsection:energy+BE2}
\begin{figure}[ht!]
	\centering\includegraphics[width=9cm]{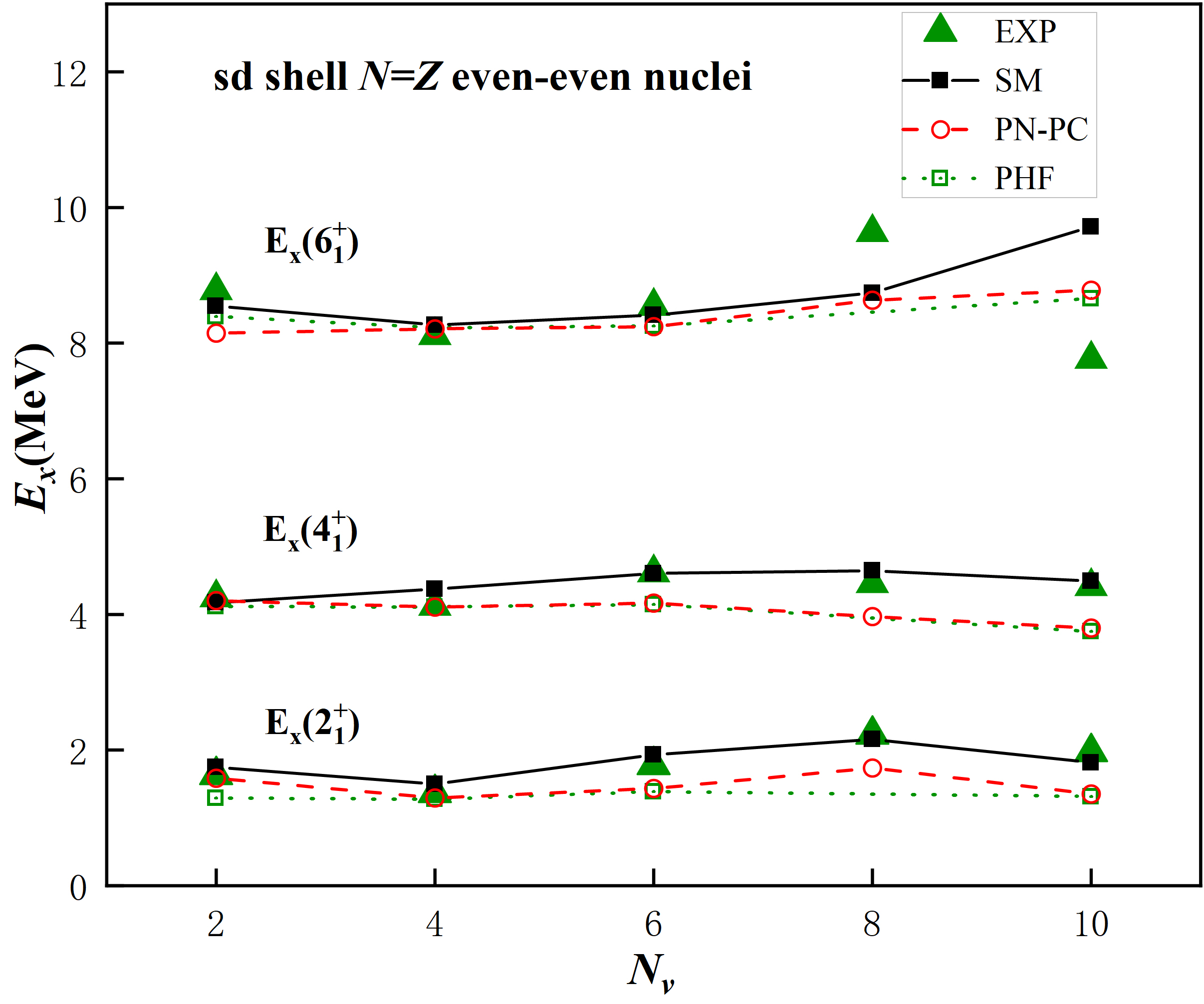}
	\caption{Excitation energy of low-lying states of even-even N=Z nuclei in sd shell. “EXP” stands for experimental data, “SM” stands for the full configuration-interaction shell model, “PN-PC” stands for angular-momentum projection after variation of a proton-neutron pair condensate (this work), and “PHF” stands for the Projected Hartree-Fock method. \label{fig:sd_even_even_N=Z}}
\end{figure}

In Fig. \ref{fig:sd_even_even_N=Z} we present excitation energies of the first $2^+, 4^+, 6^+$ states for all even-even $N=Z$ nuclei in the $1s0d$ major shell.
There are five $N=Z$ even-even nuclei between doubly magic $^{16}$O and $^{40}$Ca, with valence particles $N_\pi = N_\nu = 2,4,6,8,10$.
The USDB \cite{PhysRevC.74.034315} interactions are used in all the theoretical computations, i.e., the shell model\cite{2013Factorization} (SM), the projection after variation of one proton-neutron pair condensate (PN-PC), and the projected Hartree Fock method\cite{PhysRevC.96.064304} (PHF).
Given the valence space and the interactions, the shell model solves the many-body Schrodinger's equation by ``exact" diagonalization, therefore the shell model results are taken as a benchmark for the approximate methods.
The ground state energy from PN-PC (with one proton-neutron pair condensate as the reference state) is higher than the ``exact" ground state of the shell model, by $0.58, 1.58, 1.84, 3.75, 1.09$ MeV respectively. 
The excitation energies from PN-PC and PHF turn out to be close to shell model results, and all the theoretical methods are in qualitative agreement with experimental data.
\begin{table}[htbp]
 \centering
 \caption{B(E2) values among yrast states for even-even $N=Z$ nuclei in the $1s0d$ major shell, in units of $e^2 fm^4$. ``Exp" stands for experimental data \cite{FromENSDFdatabaseasofJulythe1st},``PN-PC" stands for proton-neutron pair condensate, and ``SM" stands for shell model. \label{tab:BE2-even-even-sd} 
 \label{tab:sd-even-even-N=Z}
 }
 \begin{tabular}{ccccccc}
  \hline  % 顶部线
  \hline
  Nucleus&Method&$2_1^{+} \rightarrow 0_1^{+}$&$4_1^{+} \rightarrow 2_1^{+}$&$6_1^{+} \rightarrow 4_1^{+}$&$8_1^{+} \rightarrow 6_1^{+}$ \\ 
  \hline  % 中部线
  $^{20}\rm Ne$ & Exp & 65.46 & 70.94 & 64.49 & 29.02 \\
  & SM    & 44.74 & 52.8  & 39.26 & 26.75 \\
  & PN-PC & 46.75 & 55.96 & 44.44 & 26.14 \\
  \hline  % 底部线
  $^{24}\rm Mg$ & Exp & 86.64 & 146.8 & 156.26 &  \\
  & SM & 73.22 & 95.92 & 88.39 & 8.05 \\
  & PN-PC & 72.09 & 96.42 & 93.89 & 80.8 \\
  \hline
  $^{28}\rm Si$ & Exp & 66.67 & 82.83 & 53.53 & \\
  & SM    & 78.48 & 110.41 & 94.71 & 68.18 \\
  & PN-PC & 87.94 & 120.59 & 120.86 & 106.77 \\
  \hline
  $^{32}\rm S$ & Exp & 71.21 & 84.49 & & \\
  & SM  & 48.14 & 68.54 & 46.13 & 35.49 \\
  & PN-PC & 38.14 & 55.95 & 55.18 & 44.22 \\
  \hline
  $^{36}\rm Ar$ & Exp & 57.9 & 84.73 & & \\
  & SM & 52.81 & 65.7 & 54.27 & 32.3 \\
  & PN-PC & 52.76 & 67.22 & 56.65 & 33.75 \\
  \hline  
 \end{tabular}
\end{table}
In Tab. \ref{tab:sd-even-even-N=Z} the B(E2) values among yrast states are presented.
Qualitative agreements are achieved among the shell model results, PN-PC results, and experimental data.

\begin{figure}[ht!]
	\centering\includegraphics[width=0.9 \textwidth]{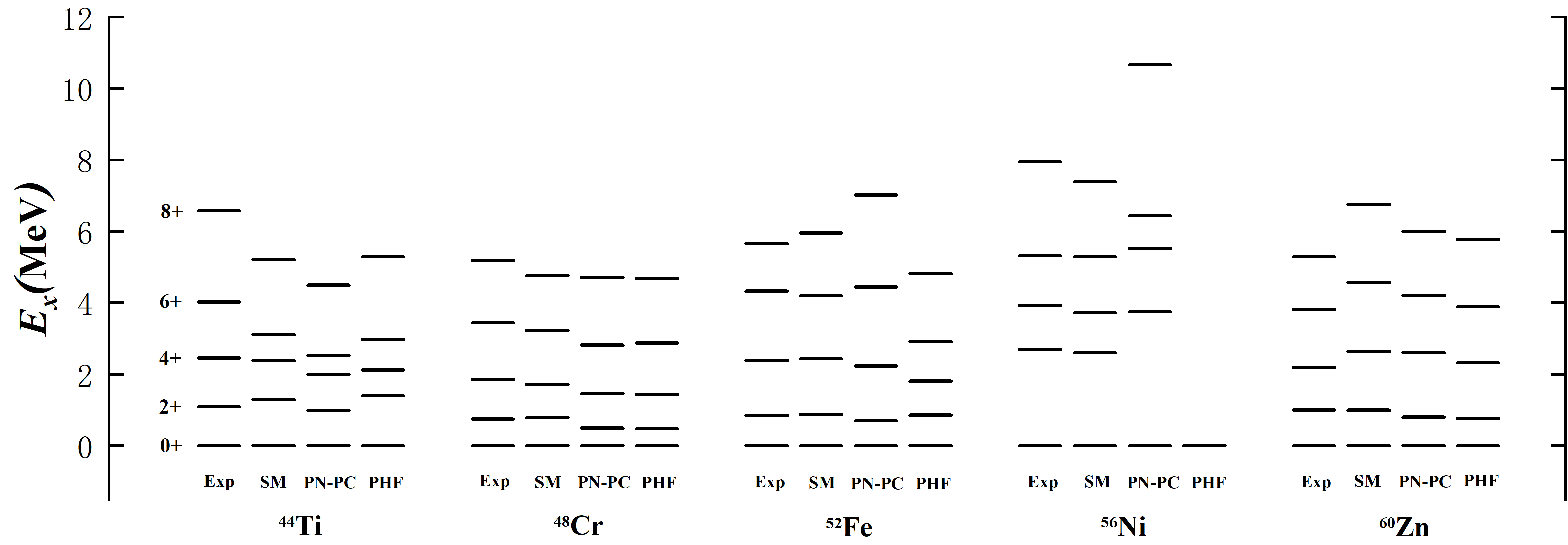}
	\caption{Ground-band excitation energy of even-even N=Z nuclei in the $1p0f$ major shell: $^{44}Ti$,$^{48}Cr$,$^{52}Fe$,$^{56}Ni$,$^{60}Zn$. \label{fig:pf-spectrum} }
\end{figure}

In Figure \ref{fig:pf-spectrum}, we present yrast states of even-even nuclei in the $1p0f$ major shell, with the GX1A \cite{Honma2005} interactions. 
For $^{44}$Ti, $^{48}$Cr, $^{52}$Fe, $^{60}$Zn, PN-PC slightly underestimate the $E_x(2^+_1)$.
For $^{56}$Ni, the Hartree-Fock method gets a spherical minimum, therefore no excited states are projected out.
Neither does PN-PC generate spectrum close to shell model results.
As $^{56}$Ni has shape-coexistence, more reference states can be needed, but in this work we are restricted to one-reference results.

%Figure \ref{fig: odd-odd-sd-pf} presents the low-lying spectra of odd-odd N=Z nuclei in sd and pf shell. For nuclei with valence proton numbers of 1, 3, and 5 in sd shell, PVPC agrees well with the low-lying spectrum of shell model and experimental data. For nuclei with valence proton numbers of 3 and 5 in pf shell, the PVPC is $0.1\sim1.2$Mev above the shell model.
%\begin{figure}[ht!]
%	\centering\includegraphics[width=16cm]{figure3.png}
%	\caption{ Low-lying spectrum of odd-odd N=Z nuclei in sd and pf shell.
%\label{fig: odd-odd-sd-pf}	
%}
%\end{figure}

\begin{figure}[ht!]
	\centering\includegraphics[width=0.7\textwidth]{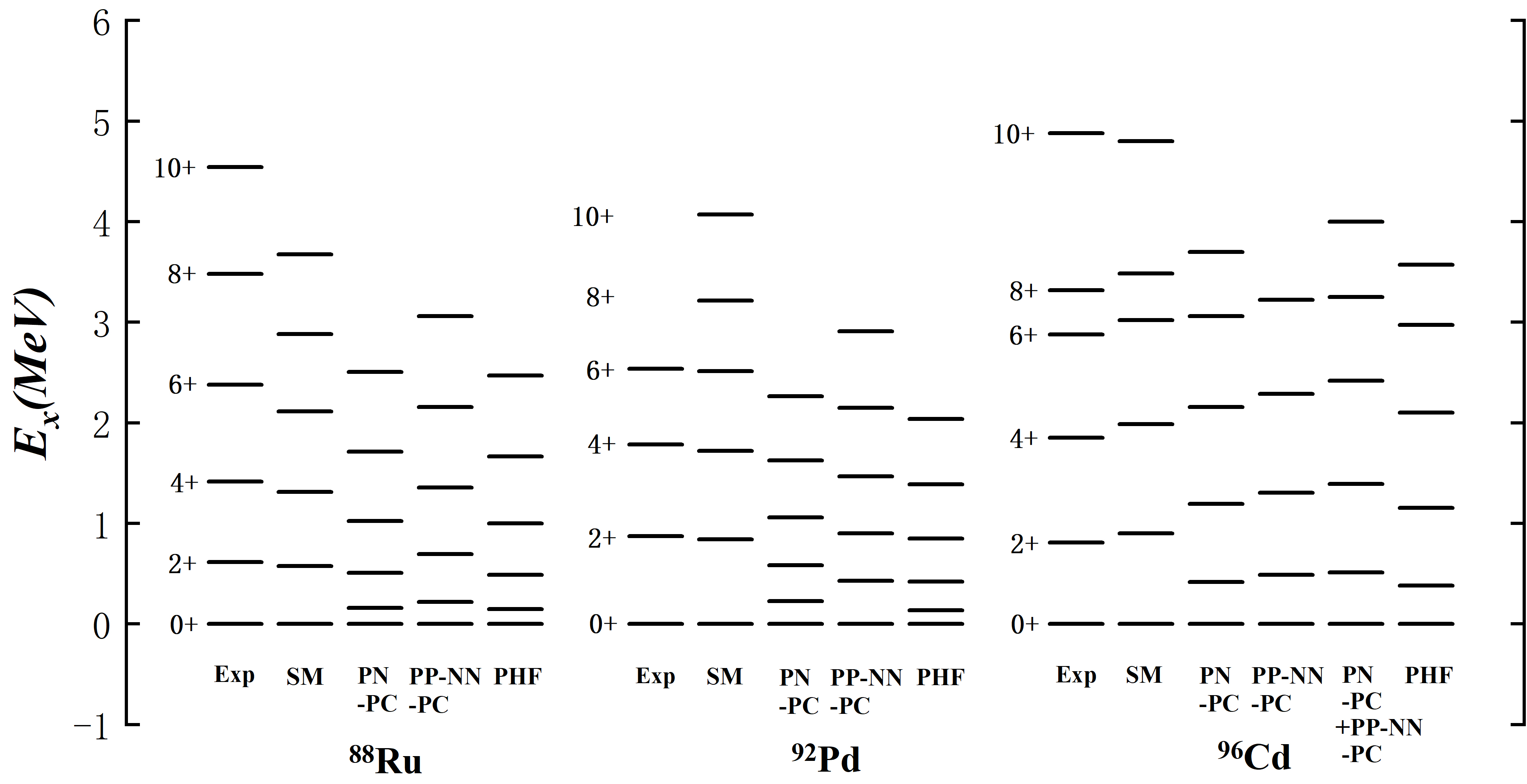}
	\caption{Excitation energy for yrast states of $^{88}$Ru, $^{92}$Pd and $^{96}$Cd. ``PN-PC” denotes yrast states projected from a proton-neutron pair condensate, ``PP-NN-PC" denotes those projected from a proton-proton neutron-neutron pair condensate, “SM” stands for the full configuration-interaction shell model, and ``EXP” stands for experimental data, ``PHF" stands for Projected Hartree-Fock states. All the theoretical computations are in the $1p_{1/2}1p_{3/2}0f_{5/2}0g_{9/2}$ major shell with JUN45 effective interactions. }
    \label{fig:88Ru-92Pd-96Cd}
\end{figure}

In Fig. \ref{fig:88Ru-92Pd-96Cd}, we present yrast bands for three even-even $N=Z$ nuclei: $^{88}$Ru, $^{92}$Pd and $^{96}$Cd.
The experimental data (Exp)\cite{Cederwall2010} and shell model results (SM)\cite{Qi2011} show a vibrational pattern for the yrast band, i.e. the $0^+, 2^+, 4^+, 6^+$ states are equally spaced approximately.
All theoretical results are computed with the JUN45 \cite{Honma2009} effective interactions, in the $1p0f_{5/2}0g_{9/2}$ major shell.
``PN-PC" denotes states projected from one proton-neutron pair condensate, while ``PP-NN-PC" denotes that of proton-proton neutron-neutron pair condensate, ``PHF" denotes states projected from one Slater determinant, i.e. a Hartree-Fock minimum.
It turns out PN-PC, PP-NN-PC and PHF underestimate the $E_x(2^+)$ gap more or less, although the gaps between $2^+, 4^+, 6^+$ are close to constant.
In all these approximation methods, only one $0^+$ states can be projected out from one reference state, but for higher-$J$ states at most $2J+1$ bases can be projected out, upon which further mixing is induced when solving the Hill-Wheeler equation.
%Therefore, for higher-$J$ states, configuration-mixing is more abundant in these models, and possibly as a result the $0^+_1$ is not low enough.
It is interesting to explore the strategy of including more reference states, so to improve the accuracy of the ground state energy.
One approach is via the Generator Coordinate Method, though in this work we are restricted to one-reference state results, except for $^{96}$Cd we introduce the mixing between PN-PC and PP-NN-PC, and the spectrum does get slightly better in Fig. \ref{fig:88Ru-92Pd-96Cd}.
Overall PP-NN-PC performs better than PN-PC, especially for $^{92}$Pd, PP-NN-PC gets closer to the shell model results.

\begin{table}[htbp]
 \centering
 \caption{B(E2) values for $^{88}$Ru, $^{92}$Pd and $^{96}$Cd, in units of $e^2 fm^4$. ``PN-PC" stands for proton-neutron pair condensate, ``PP-NN-PC" stands for proton-proton neutron-neutron pair condensate, and ``SM" stands for shell model. \label{tab:BE2-88Ru-92Pd-96Cd} }
 \begin{tabular}{ccccccc}
  \hline  % 顶部线
  \hline
  Nucleus&Method&$2_1^{+} \rightarrow 0_1^{+}$&$4_1^{+} \rightarrow 2_1^{+}$&$6_1^{+} \rightarrow 4_1^{+}$&$8_1^{+} \rightarrow 6_1^{+}$&$10_1^{+} \rightarrow 8_1^{+}$ \\ 
  \hline  % 中部线
  $^{88}\rm Ru$&SM&445.79&693.11&803.60&886.38&959.59 \\
  &PN-PC&659.32&936.18&1019.18&1047.02&934.73 \\
  &PP-NN-PC&624.36&887.28&966.94&995.14&997.97 \\
  $^{92}\rm Pd$&SM&276.17&347.15&330.81&286.01&304.15\\
  &PN-PC&216.7&304.71&325.66&325.26&312.79 \\
  &PP-NN-PC&210.66&298.63&318.60&317.36&306.01 \\
  $^{96}\rm Cd$&SM&138.57&187.85&174.20&42.59&47.41 \\
  &PN-PC&141.79&194.24&195.07&174.78&142.14 \\
  &PP-NN-PC&134.28&189.12&193.07&174.19& 143.06 \\
  \hline  % 底部线
  \hline  
 \end{tabular}
\end{table}
While the vibrator limit may indicate that the $B(E2, I \rightarrow I-2)$ value increases linearly as the angular momentum $I$ of the yrast state increases\cite{Qi2011}, shell model B(E2) values of these three nuclei are roughly constant.
In Tab. \ref{tab:BE2-88Ru-92Pd-96Cd} we present B(E2) values from the shell model, PN-PC and PP-NN-PC.
The effective charges are taken as $e_p = 1.5e, e_n = 0.5e$, and the harmonic oscillator length is taken simply as $r_0 = A^{1/6}fm$.
Both PN-PC and PP-NN-PC generate B(E2) values qualitatively close to shell model results, except for B(E2, $8^+_1 \rightarrow 6^+_1$) and B(E2, $10^+_1 \rightarrow 8^+_1$) in $^{96}$Cd, albeit experimental B(E2) values are not available yet, as far as we know.
Consistent with the excitation spectrum in Fig. \ref{fig:88Ru-92Pd-96Cd}, the PP-NN-PC results of B(E2) values are slightly better than the PN-PC.

\begin{figure}[ht!]
\centering\includegraphics[width=16cm]{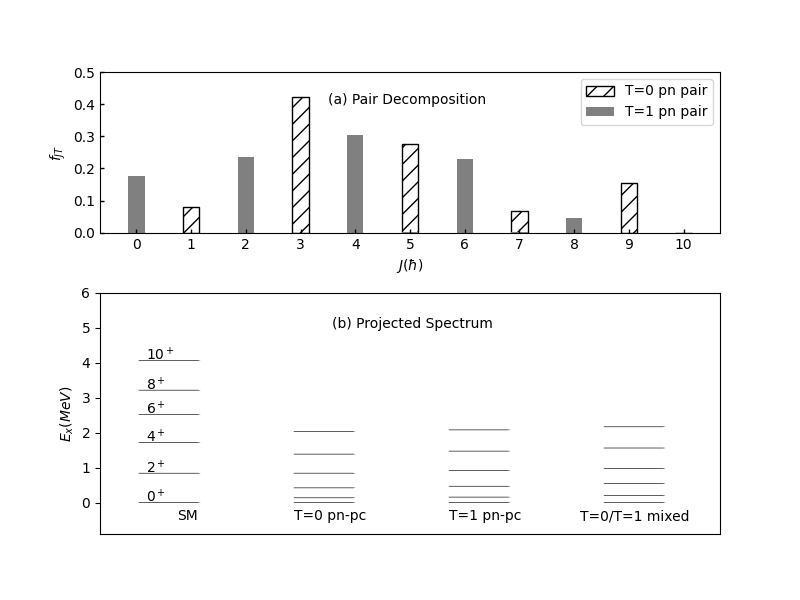}
\caption{ 
States of $^{92}$Pd. ``T=0 PN-PC" denotes states projected out of a T=0-proton-neutron-pair condensate, ``T=1 PN-PC" denotes that of a T=1-proton-neutron-pair condensate, ``T=0/T=1 mixed" denotes the states with configuration mixing among the T=0 PN-PC and T=0 PN-PC. After variation, the optimized T=0 proton-neutron pair turns out to consist of only $J$=odd components, while the T=1 proton-neutron pair only have $J$=even components. However when they construct condensates respectively, the projected spectrum are almost the same.
\label{fig: 92pd_dual_description}	
}
\end{figure}

Ref. \cite{Zerguine2011,Neergaard2013,Fu2013,Fu2016} discussed the dual description of pair coupling schemes in these $N=Z$ even-even nuclei.
Traditionally, strategies of like-nucleon pair truncation favor low-angular-momentum T=1 pairs, such as S, D, G, I pairs for $J=0,2,4,6$ respectively, while $\delta$-interaction analysis shows $T=0, J=J_{max}/J_{min}$ proton-neutron pairs can be important \cite{ZHAO20141}.
As the angular momentum coupling scheme of pairs is not unique, Ref. \cite{Zerguine2011,Neergaard2013,Fu2013,Fu2016} argues that, for $N=Z$ even-even nuclei, these two schemes may both describe the low-lying states well, constructing a dual description.
In our framework, we address this issue by restricting the proton-neutron pair to be T=0 or T=1 respectively, and see the outcome of the proton-neutron pair condensate.
In other words, we optimize condensates of T=0 and T=1 pn-pair condensates respectively,
\begin{eqnarray}
\left| (\hat{A}^{\dagger}_{\pi\nu, T=0} )^{N_\pi} \right\rangle, ~~~~~~~
\left| (\hat{A}^{\dagger}_{\pi\nu, T=1} )^{N_\pi} \right\rangle.
\end{eqnarray}
In Fig. \ref{fig: 92pd_dual_description} panel (a), we present the decomposition of the optimized T=0 and T=1 proton-neutron pair respectively.
It turns out, after variation, the T=0 proton-neutron pair consists of only $J$=odd components, while the T=1 proton-neutron pair consists of only $J$=even components.
In Fig. \ref{fig: 92pd_dual_description} panel (b), we present the projected spectrum in excitation energies.
The T=0 PN-PC and the T=1 PN-PC have almost the same projected spectra, and it does not improve much by mixing them up.
Therefore a dual description occurs also in such a one-reference benchmark.
PN-PC assumes the trial wavefunction is a duplication of the same proton-neutron pair, and optimizes the inner structure of the proton-neutron pair.
PP-NN-PC assumes the trial wavefunction is a direct product of one proton pair condensate and one neutron pair condensate, allowing the proton pair and neutron pair to differ in structure.
With the dual description said, the PP-NN-PC trial wavefunction may reserve more generality, which may explain why PP-NN-PC outperforms PN-PC moderately in the yrast spectrum in Fig. \ref{fig:88Ru-92Pd-96Cd}.

%In this section we focus on the entropy of optimized pair condensates.
\subsection{ Shannon entropy for $^{18}$F $\sim$ $^{98}$In } \label{subsec:shannon-entropy}
As defined in Eq. (\ref{eqn:entropy}), the Shannon entropy measures the probability distribution among proton-neutron pair configurations on canonical bases.
The more even the probability distribution is, the larger the entropy becomes.
\begin{figure}[h!]
\includegraphics[width=0.8\linewidth]{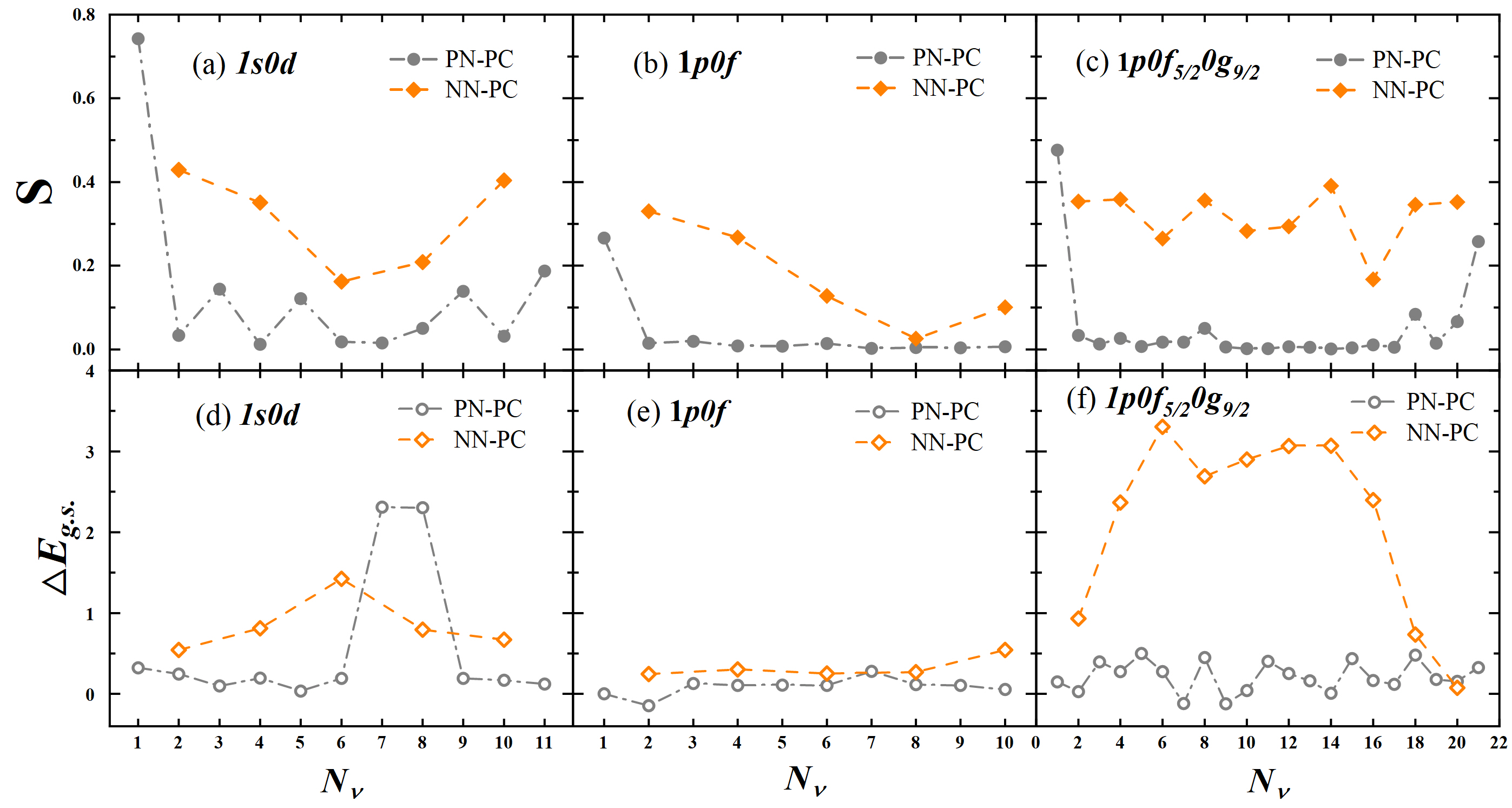}
\caption{
Panel (a): Entropy of optimized proton-neutron pair condensates for $1s0d$-shell nuclei from $^{18}$F to $^{38}$K, in comparison with semi-magic oxygen isotopes.
``PN-PC" denotes proton-neutron pair condensates, while ``NN-PC" denotes neutron-neutron pair condensates; all those condensates are optimized by variation with shell model interactions.
%As the entropy of optimized proton-neutron pair condensates is small, logarithmic vertical coordinates are taken.
Panel (b): Entropy for $1p0f$-shell $N=Z$ nuclei from $^{42}$Sc to $^{60}$Zn, semi-magic nuclei $^{42}$Ca to $^{50}$Ca.
Panel (c): Entropy for $1p0f_{5/2}0g_{9/2}$-shell $N=Z$ nuclei from $^{58}$Cu to $^{98}$In, semi-magic nuclei $^{58}$Ni to $^{76}$Ni.
\label{fig:entropy}
}
\end{figure}
In Fig. \ref{fig:entropy} Panel (a-c) we present entropies of the optimized proton-neutron pair condensates for all $N=Z$ nuclei from $^{18}$F to $^{98}$In.
Correspondingly, we show in Fig. \ref{fig:entropy} Panel (d-f) the extra energetic descent in the ground state generated by pair condensates,
\begin{equation}
\Delta E_{g.s.} \equiv E^{PHF}_{g.s.} - E^{PVPC}_{g.s.}.
\end{equation}
$E^{PVPC}_{g.s.}$ is the ground state projected out of the variationally optimized pair condensate, while $E^{PHF}_{g.s.}$ is the ground state projected out of the Hartree-Fock minimum.
Because of extra pairing correlation in the pair condensate trial wavefunction, typically $E^{PVPC}_{g.s.}$ is energetically lower than $E^{PHF}_{g.s.}$.
Therefore $\Delta E_{g.s.}$ defines the extra energetic descent in the ground state, due to the more generality of the pair condensate ansatz.
When the entropy increases, $\Delta E_{g.s.}$ is expected to be increase too, though these two quantities are not in proportion.

The nuclei from $^{18}$F to $^{38}$K are computed with the USDB interactions, those from $^{42}Sc$ to $^{60}$Zn are computed with the GX1A interactions, and those from $^{58}$Cu to $^{98}$In are computed with the JUN45 interactions.
For five cases of one valence proton (hole) and one valence neutron (hole) outside a doubly magic core, $^{18}$F, $^{38}$K, $^{42}$Sc, $^{58}$Cu, $^{98}$In, a proton-neutron condensate generates the shell-model ground state exactly, and ends up with significant entropy beyond $S=0.2$ in Panel (a-c).
However, when one additional proton-neutron pair is added into the deuteron-like condensate, the entropy drastically decreases below 0.1, which means the proton-neutron pair condensate is not far from a Slater determinant mean-field solution.
$\Delta E_{g.s.}$ for PN-PC is below 0.5 MeV in Panel (d-f), in support of this conclusion.

In comparison, the entropy of semi-magic oxygen, calcium, nickel isotopes are almost all notably above 0.1, signaling a correlated neutron-neutron condensate for those nuclei.
Therefore, the NN-PC results align with the conventional wisdom that there is a correlated, or superfluidic pair condensate in semi-magic nuclei.
For the semi-magic nuclei, the error of the ground state energy of PVPC is about $0.5$ MeV, close to exact ground state of the shell model, in support for the T=1 correlated pair condensate ``phase".
The PN-PC ground states are typically 1-2 MeV above the shell-model ground states, less accurate than the NN-PC.
Considering such a deviation of PN-PC against shell-model, we note that these one-reference results may not be conclusive enough, albeit the absence of an entangled PN-PC hints negatively against the picture of an entangled proton-neutron pair condensate ``phase".

It is notable that odd-even staggering of the PN-PC entropy is observed in the $1s0d$ major shell, which may imply there can be more proton-neutron correlations in odd-odd $N=Z$ nuclei beyond mean field, and appears in alignment with a recent study of shell-model-alike approach based on relativistic density functional theory\cite{Wang2024}.
However, in our calculations, such an odd-even staggering becomes negligible in $1p0f_{5/2}0g_{9/2}$ shell, implying possible dependence on the valence space or effective interactions used.
The $N=28$ subshell effect is shown in Fig. \ref{fig:entropy}(b).
The entropy of neutron-neutron pair condensate reduces in Calcium isotopes as the valence neutron number $N_n$ approaches $28$, gets a nonzero minimum at $^{48}$Ca, and then rises again in $^{50}$Ca.

\subsection{Competition between T=1 and T=0 pair condensates} \label{subsec:t1t0-competition}
It is interesting to consider the competition between T=1 and T=0 configurations in nuclear wavefunctions, as a result of T=1 and T=0 forces in the Hamiltonian.
Shell model wavefunctions in the $1p0f$ shell have been used to analyze the possibility of T=0 proton-neutron pair condensate\cite{Poves1998}, with the elegant Lanczos algorithm applied to shell model codes\cite{RevModPhys.77.427}. 
The shell model ground state with realistic interactions are considered as physical solutions\cite{Poves1998}, while the ground state with T=0/T=1 interactions turned off are considered as T=1/T=0 ``pair condensate", to an approximation.
It was found the overlap between the physical solution and the T=0 ``condensate" is $16\%$ for $^{44}$Ti, and $2\%$ for $^{48}$Cr.
Pairing phase transition in $^{24}$Mg and $^{28}$Si was studied with shell model wavefunctions, regarding expectation values of pair counting operators as a function of pairing strength, excitation energy, angular momentum and different types of interactions \cite{Horoi2007}. 
Pair correlation energies are also studied as a function of temperature and angular momentum\cite{Sheikh2008}.

With the general proton-neutron pair condensate as the ansatz wavefunction, we revisit this numerical experiment in the lens of Shannon correlation entropy, so as to investigate the impact of interaction forces on the entropy, and possible ``phase" transitions driven by parameters.
Given the USDB interactions which has been successful in describing a wide range of experimental data \cite{PhysRevC.74.034315}, we multiply all the T=1 interactions with $(1-x)$, and the T=0 interactions $(1+x)$,
\begin{equation}
\hat{H}' = \hat{H}^{USDB}_0 + (1-x)\hat{H}^{USDB}_{T=1} + (1+x)\hat{H}^{USDB}_{T=0}, \label{eqn:T1vsT0}
\end{equation}
where $\hat{H}^{USDB}_0$ denotes the one-body parts of the USDB interactions, $\hat{H}^{USDB}_{T=1}$ is the T=1 two-body parts, and $\hat{H}^{USDB}_{T=0}$ is the T=0 two-body parts.
When $x=-1$, there is only T=1 interactions (doubled); when $x=1$, there is only T=0 interactions (doubled); as $x$ varies from $[-1,1]$ the Hamiltonian transits artificially from a T=1 limit to T=0 limit.
We note that only $x=0$ is the physical situation, but such a numerical experiment reveals how the model outcome responds to external parameters.
We choose $^{24}$Mg in the $1s0d$ shell as an example.
With different $x$ values, we do variation of a general proton-neutron pair condensate, and measures its Shannon correlation entropy.
As comparison, we also do variation of a proton-proton neutron-neutron pair condensates as in Ref. \cite{PhysRevC.105.034317}.
We denote ``PN-PC" as the proton-neutron pair condensate, in which the proton-neutron pair can be mixed with T=0 and T=1 components.
We denote ``PP-NN-PC" as the proton-proton neutron-neutron pair condensate, in which the like-nucleon pairs can only have T=1.

\begin{figure}[h!]
\includegraphics[width=0.7\linewidth]{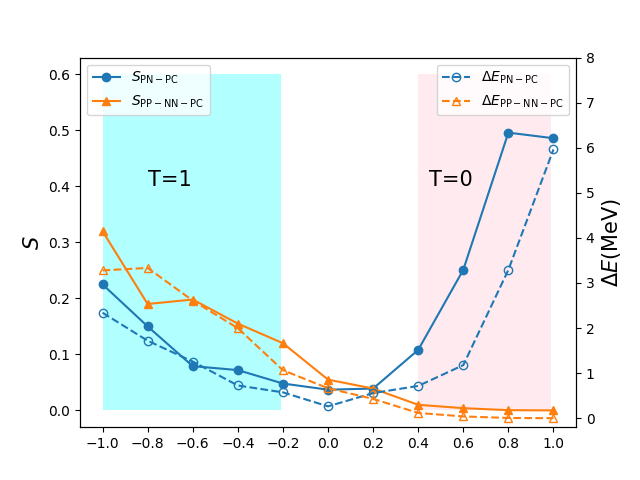}
\caption{
Entropy of optimized pair condensates for $^{24}$Mg in the $1s0d$ major shell, under artificial Hamiltonian $\hat{H} = \hat{H}^{USDB}_0 + (1+x)\hat{H}^{USDB}_{T=0} + (1-x)\hat{H}^{USDB}_{T=1}$, with $x \in [-1,1]$. ``PN-PC" stands for proton-neutron pair condensate, while ``PP-NN-PC" stands for proton-proton neutron-neutron pair condensate. When $x=0$ the physical situation has low correlation entropy for both PN-PC and PP-NN-PC; when $x$ goes away from $0$, the pair condensates onset T=0 and T=1 ``phase" transitions. The corresponding energetic descent of the pair condensates due to correlation further than a Hartree-Fock solution is denoted as $\Delta E_{\rm PN-PC}$ and $\Delta E_{\rm PP-NN-PC}$ respectively.
\label{fig:mg24_entropy_T1T0}
}
\end{figure}

In Fig.\ref{fig:mg24_entropy_T1T0} we show the entropies of pair condensates for $^{24}$Mg, optimized with the Hamiltonian (\ref{eqn:T1vsT0}).
The solid circles and solid triangles in Fig. \ref{fig:mg24_entropy_T1T0} are for the entropies of PN-PC and PP-NN-PC respectively.
At $x =0$, both the PN-PC and PP-NN-PC have very small entropy around 0.05, implying the condensates are not quite far from a Slater determinant, i.e. the Hartree-Fock minimum.
But as $x$ increases in $x \in [0, 1]$, the PN-PC has larger entropy while the PP-NN-PC has even smaller entropies approaching absolute zero.
This signals the onset of T=0 correlated condensate ``phase" transition.
At $x=0.8$ the entropy of PN-PC reaches maximum of about 0.5, and saturates there for $x=1.0$.
As $x$ decreases from zero to $-1$, both PN-PC and PP-NN-PC have larger entropies, approaching 0.2 $\sim$ 0.3 at $x = -1$.
As the T=1 interactions are symmetric for proton-proton, neutron-neutron, proton-neutron pairs, and our proton-neutron pair is general without restricting its isospin, both PN-PC and PP-NN-PC transit to a T=1 condensate ``phase".
The proton-neutron pair configurations are richer with both T=1 and T=0 channels, therefore the $d$ value in the definition of Shannon correlation entropy (\ref{eqn:entropy}) is larger for PN-PC, resulting in smaller entropies, compared with 0.5 at the T=0 limit with $x=0.8, 1.0$.
If we set $S=0.1$ as the threshold to label the onset of a correlated condensate ``phase", these results imply one needs about $x=0.4$ to get a proton-neutron correlated condensate phase.

Corresponding to the entropy, we also compare the energy descent of a pair condensate against a Hartree-Fock solution.
Under the same Hamiltonian (\ref{eqn:T1vsT0}), we perform Hartree-Fock calculations with the codes $Sherpa$ \cite{PhysRevC.66.034301,Johnsonb}.
As a pair condensate is a more general ansatz wavefunction, its energetic minimum is lower than that of the Hartree-Fock solution.
We denote their energetic difference
\begin{eqnarray}
\Delta E_{\rm PN-PC} &=& E^{\rm HF}_{min} - E^{\rm PN-PC}_{min}, \\
\Delta E_{\rm PP-NN-PC} &=& E^{\rm HF}_{min} - E^{\rm PP-NN-PC}_{min}.
\end{eqnarray}
as signals of the gain from extra correlations in the pair condensates.
Note that $E_{min}$ denotes the energetic minimum after the variation, without further angular momentum projection yet.
In Fig. \ref{fig:mg24_entropy_T1T0} the void circles and void triangles stand for $\Delta E_{\rm PN-PC}$ and $\Delta E_{\rm PP-NN-PC}$ respectively.

Both the entropy $S$ and the energetic descent $\Delta E$ are diagnostic signals of pair condensates based on comparison against Hartree-Fock solutions.
It is shown that $S$ and $\Delta E$ are highly consistent.
When $x=0$ in the physical situation, the entropies and energies of PN-PC and PP-NN-PC are quite close, thus they construct dual descriptions.
The PP-NN-PC energy descends a bit deeper, so the energy gain $\Delta E_{\rm PP-NN-PC}$ is larger than the PN-PC, though both of them have small entropies.
%In this sense, we might say the PP-NN-PC has a narrow victory against PN-PC in $^{24}$Mg.

\begin{figure}[h!]
\includegraphics[width=1.0\linewidth]{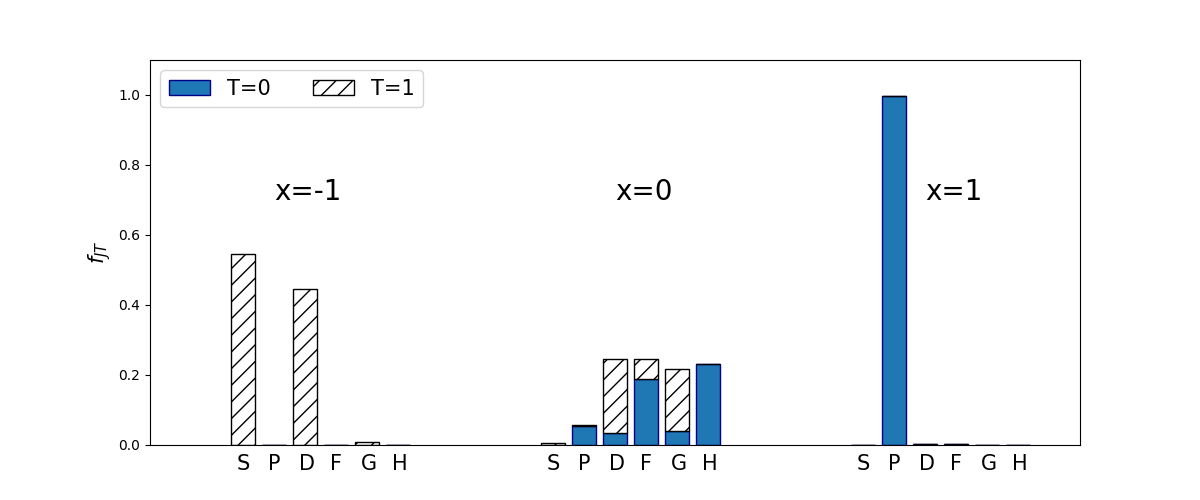}
\caption{
Fractions of different pairs with $J=0,\cdots, 5$ and $T=0,1$ in the optimized pair of the proton-neutron pair condensate for $^{24}$Mg, given the artificial Hamiltonian in (\ref{eqn:T1vsT0}).
When $x=0$ the physical solution has all kinds of pairs of T=0 and T=1 mixed up; when $x=-1$, T=1 S and D pairs dominate; when $x=1$ the T=0 P pairs dominate. 
S, P, D, F, G, H stand for $J=0,1,2,3,4,5$ pairs respectively.
\label{fig:mg24_fJ}
}
\end{figure}

\subsection{Anatomy of the pairs}

To look closer into the microscopic details of the proton-neutron pair condensates, we decompose them into components with good angular momentum and isospin, $\hat{A}^\dagger_{\pi \nu} = \sum_{JMT} \hat{\mathbbm A}^{JMT\dagger}_{\pi \nu}$, as mentioned in Sec. \ref{subsec:variation}.
Norms of such pairs can be easily computed with generalized Wick's theorem \cite{JinQuan1993}, and the fraction of $(J,T)$ components in a given proton-neutron pair, $f_{JT}$, is defined in (\ref{eqn:fJT}).

In $1s0d$ shell the angular momentum of a pair can only be $J = (0,1,2,3,4,5)$, we label pairs with such angular momenta as (S,P,D,F,G,H) pairs.
In Fig. \ref{fig:mg24_fJ}, we show the anatomy of the proton-neutron pairs studied in Fig. \ref{fig:mg24_entropy_T1T0}, with $x= 0, \pm 1$, corresponding to physical interactions and artificial interactions in the T=1, T=0 limits.
When the external parameter in (\ref{eqn:T1vsT0}), $x=-1$, i.e. in the T=1 limit, the optimized proton-neutron pair condensate is made of T=1 S, D pairs.
This is in alignment with the conventional textbook knowledge that T=1 low-angular-momentum pairs are important \cite{Casten1990}.
In the physical situation of $x=0$, the optimized proton-neutron pair condensate is a mixture of both T=1 and T=0 pairs.
The stacked bars in Fig. \ref{fig:mg24_fJ} mean that both T=0 and T=1 pairs with a given $J$ can coexist, e.g. there are both T=0 and T=1 D, F, G pairs when $x=0$.
In the T=0 limit, the fraction of T=0 P pair increases, and there onsets a pure P-pair condensate \cite{VanIsacker2016}.
This also reveals that, the T=0 components under physical interactions are quite different from those under purely T=0 interactions\cite{Poves1998}.
Only when the T=0 interactions dominates, does the PN-PC becomes a highly-entangled (with large entropy), P-pair-dominating condensate.

With angular momentum projection techniques, we find that four proton-neutron P-pairs in $^{24}$Mg are in alignment, coupling into total angular momentum $J=4$.
Similar numerical experiments show that, in the T=0 limit of the modified USDB and GX1A interactions respectively, $^{32}$S, $^{48}$Cr, $^{72}$Kr all end up with a purely P-pair condensate.
However, curiously, in the $1p0f_{5/2}0g_{9/2}$ with the JUN45 interactions, in the T=0 limit, $^{64}$Ge ends up a P-pair condensate, but $^{92}$Pd ends up a $J=9$ pair condensate.

\section{Summary and Outlook}\label{summary}

In summary, we have implemented angular momentum projection after variation of an explicit general proton-neutron pair condensate (PN-PC), with good particle numbers, under shell model effective interactions.
With one optimized proton-neutron pair condensate as the reference state, we benchmark PN-PC against shell model results, the Projected Hartree-Fock method, and experimental data when available.
Qualitative agreements are achieved on yrast excitation energies and B(E2) values for a series of even-even $N=Z$ nuclei in the $1s0d$ and $1p0f$ major shell, though the PN-PC one-reference ground state energies are around 1-2 MeV above that of the shell model.
In the $1p0f_{5/2}0g_{9/2}$ major shell, one unique proton-neutron pair condensate does not reproduce the vibrational-type yrast spectrum of $^{88}$Ru, $^{92}$Pd and $^{96}$Cd, but qualitatively reproduces the B(E2) values among yrast states.
A dual description is observed in $^{92}$Pd, i.e. the optimized T=0 proton-neutron condensate and the optimized T=1 proton-neutron condensate generate similar spectrum and B(E2) values.
Analysis of Shannon correlation entropy reveals that, like-nucleon pair condensates in semi-magic nuclei show an entangled phase, while optimized proton-neutron pair condensates in $N=Z$ nuclei from $^{18}$F to $^{98}$In are not far from a mean-field Slater determinant, as a result of the complicity of both T=0 and T=1 interactions.

With artificial interactions varying T=1 and T=0 strengths, we show that the proton-neutron pair condensate for $^{24}$Mg transits from a T=1 entangled phase to a T=0 entangled phase. 
Both the T=1 limit and the T=0 limit have large entropies, with typical $(J,T)$ anatomy of the optimized proton-neutron pair.
In the T=1 limit the optimized proton-neutron pair consists of T=1 S, D pairs primarily; and in the T=0 limit the optimized proton-neutron pair is a pure T=0 P pair condensate.
However with physical interactions (USDB), the optimized proton-neutron pair can be a mixture of both T=1 and T=0 pairs.
As the T=0 pairing strength increases, the fractions of high-$J$ pairs decrease and the fraction of T=0 P pair increases.

All results presented in this work are based upon one optimized proton-neutron pair condensate, i.e. ``one-reference" results.
In future work, we expect the results can be improved by including more reference states, as in the Generator Coordinate Method.
It is also interesting to look more carefully at condensates with broken pairs, i.e. allow the ansatz wavefunction to have impurity pairs, rather than restricting it to be a condensate of the same pairs.
And it is interesting, though challenging, to apply variation after projection techniques, as in Ref. \cite{Gao2022} and references therein, to proton-neutron pair condensates, so that more details may be revealed for each of the yrast state in $N=Z$ nuclei.

\acknowledgements

We would like to thank discussions with Xin Guan on Shannon entropy of pairing configurations.
Yi Lu acknowledges support by the Natural Science Foundation of Shandong Province, China (ZR2022MA050), the National Natural Science Foundation of China (11705100, 12175115).
Yang Lei is grateful for the financial support of the Sichuan Science and Technology Program (Grant No. 2019JDRC0017), and the Doctoral Program of Southwest University of Science and Technology (Grant No. 18zx7147). 
This material is also based upon work (Calvin W. Johnson) supported by the U.S. Department of Energy, Office of Science, Office of Nuclear Physics, under Award Number  DE-FG02-03ER41272.
Guan-Jian Fu acknowledges the National Natural Science Foundation of China (12075169, 12322506), the Fundamental Research Funds for Central Universities (22120240207).

\appendix
\section{Canonical transformation of a general pair \label{canonical-transformation}
}
To do the canonical transformation of a general pair in Eq. (\ref{canonical-basis}), is to implement the matrix transformation $A = U X U^\top$, with given skew  $n\times n$ matrix $A$ \cite{Zumino1962,Gantmacher1998}.
For self-containment we brief this transformation in our practice here.
We are restricted to real pair structure coefficients, therefore $A$ is a real skew matrix.
Multiplying $A$ with imaginary unit $i$, we get a Hermitian matrix $H = iA$, which can be always diagonalized with real eigenvalues $\lambda$, and (in general) complex eigenvector $\vec{z}$,
\begin{equation}
iA \vec{z} = \lambda \vec{z},
\end{equation}
therefore $-i\lambda$ , which is $0$ or non-zero pure imaginary number, is an eigenvalue of $A$.
When $A$ is real, from the complex conjugate of the upper equation, one sees that $i\lambda$ is also an eigenvalue of $A$.
Therefore, eigenvalues of $A$ are either paired imaginary numbers $\pm i \lambda$, or $0$.
Such eigenvalues can be easily computed using subroutines in the GNU scientific library (GSL) \cite{gsl}.

For a pair of nonzero eigenvalues $\pm i\lambda$, the corresponding eigenvectors $\vec{z}^*, \vec{z}$ can be used to construct bases,
\begin{equation}
\vec{x} = \frac{ \vec{z} + \vec{z}^* }{\sqrt{2}}, ~~~
\vec{y} = \frac{ \vec{z} - \vec{z}^* }{\sqrt{2i}}.
\end{equation} 
All paired nonzero eigenvectors give
\begin{equation}
\{  \vec{x}_1, \vec{y}_1, \vec{x}_2, \vec{y}_2, \cdots, \vec{x}_k, \vec{y}_k  \}.
\end{equation}
These vectors are all orthonormal.
If these vectors do not exhaust the $R^n$ linear space, i.e. $2k < n$, then complement more orthonormal vectors $\omega_{2k+1}, \cdots, \omega_n$, so to construct an orthonormal matrix,
\begin{equation}
U = [ \vec{x}_1, \vec{y}_1, \cdots, \vec{x}_k, \vec{y}_k, \vec{\omega}_{2k+1}, \cdots \vec{\omega}_n].
\end{equation}
Because $\forall i \in [1,k], j \in[2k+1,n]$,
\begin{equation}
A \vec{x}_i = \lambda_i \vec{y}_i, ~~ A\vec{y}_i = -\lambda_i \vec{x}_i, ~~~ A\vec{\omega}_j = 0,
\end{equation}
therefore $A$ can be transformed into canonical form,
\begin{equation}
A U = U X, ~~~ X=diag\left\{ [ \begin{smallmatrix}
0 & -\lambda_1 \\ \lambda_1 & 0 
\end{smallmatrix}] ,  
[ \begin{smallmatrix}
0 & -\lambda_2 \\ \lambda_2 & 0 
\end{smallmatrix}] ,
\cdots,
[ \begin{smallmatrix}
0 & -\lambda_k \\ \lambda_k & 0 
\end{smallmatrix}] ,
0, \cdots, 0
\right\}.
\end{equation}
Make sure $\lambda_i < 0$, and denote $\nu_i \equiv -\lambda_i$, we arrive at (\ref{A=UXUtop}).

For a proton-neutron pair, e.g. in the $sd$ shell, we denote 12 proton orbits and 12 neutron orbits, define the proton-neutron pair structure coefficient matrix $A$ as a $24 \times 24$ matrix, in the form of
\begin{equation}
A = \begin{pmatrix}
0 & \kappa \\
-\kappa & 0
\end{pmatrix}.
\end{equation}
With $\vec{x} = (\begin{smallmatrix}
\vec{x}_1 \\
\vec{x}_2
\end{smallmatrix})$, to demand $A \vec{x} = \lambda \vec{y}, ~~ A\vec{y} - \lambda \vec{x}$, is to demand
\begin{equation}
\kappa^2 \vec{x}_1 = \lambda^2 \vec{x}_1, ~~~ \kappa^2 \vec{x}_2 = \lambda^2 \vec{x}_2.
\end{equation}
Let $\vec{x}_1$ be one eigenvector of $\kappa^2$, and let $\vec{x}_2 = 0$, we end up with
\begin{equation}
\vec{x} = \begin{pmatrix}
\vec{x}_1 \\
0
\end{pmatrix}, ~~~
\vec{y} = \frac{1}{\lambda} \begin{pmatrix}
0 \\
-\kappa \vec{x}_1
\end{pmatrix}.
\end{equation}
Therefore, given a general proton-neutron pair, it can be transformed into canonical form (\ref{A=UXUtop}), with proton canonical orbits and neutron canonical orbits, respectively, on the basis of which the entropy can be calculated as defined in (\ref{eqn:entropy}).
	
\bibliographystyle{apsrev4-1}
\bibliography{ref-papers}
\end{document}